\documentclass{llncs}

\usepackage{etex} 

\usepackage{mathptmx}
\usepackage[scaled=.92]{helvet}
\usepackage{courier}

\usepackage{amssymb}
\usepackage{amsmath}
\usepackage{url}

\usepackage{xspace}
\usepackage[all]{xy}
\usepackage{hyperref}
\usepackage{color}
\usepackage{bussproofs}

\renewcommand{\tt}{\ttfamily}
\newcommand{\codefont}{\small\tt}
\newcommand{\code}[1]{\mbox{\codefont{#1}}}
\newcommand{\ccode}[1]{``\code{#1}''}
\newcommand{\us}{\raise-.8ex\hbox{-}}
\newcommand{\funset}{\ensuremath{_S}} 
\newcommand{\seq}[1]{\overline{#1}} 

\newcommand{\cupdot}{\mathbin{\mathaccent\cdot\cup}} 
\newcommand{\rulename}[1]{\mbox{\sf #1}}
\newcommand{\arulename}[1]{\mbox{\it #1}} 
\newcommand{\emptyheap}{[]}
\newcommand{\pre}[1]{#1\code{'pre}}  
\newcommand{\post}[1]{#1\code{'post}} 
\newcommand{\spec}[1]{#1\code{'spec}} 
\newcommand{\ajudge}[6]{\ensuremath{#1:{#2}\mid{#3}\leftarrow{#4}\,\Downarrow\,{#6}}}
\newcommand{\acjudge}[5]{\ensuremath{#1:{#2}\mid{#3}\leftarrow{#4}\,\Downarrow\,{#5}}}

\usepackage{listings}
\lstnewenvironment{curry}{\lstset{literate={->}{{$\rightarrow{}\!\!\!$}}3}}{}
\lstnewenvironment{prolog}{}{}
\newcommand{\listline}{\vrule width0pt depth1.5ex}

\begin{document}
\pagestyle{plain}
\sloppy

\title{Combining Static and Dynamic Contract Checking\\ for Curry}

\author{Michael Hanus}
\institute{
Institut f\"ur Informatik, CAU Kiel, D-24098 Kiel, Germany \\
\email{mh@informatik.uni-kiel.de}
}

\maketitle

\begin{abstract}
Static type systems are usually not sufficient to express
all requirements on function calls.
Hence, contracts with pre- and postconditions can be used
to express more complex constraints on operations.
Contracts can be checked at run time to ensure that operations
are only invoked with reasonable arguments and return intended results.
Although such dynamic contract checking provides more reliable
program execution, it requires execution time and could lead to
program crashes that might be detected with more advanced methods
at compile time.
To improve this situation for declarative languages,
we present an approach to combine static and dynamic contract
checking for the functional logic language Curry.
Based on a formal model of contract checking for functional logic programming,
we propose an automatic method to verify contracts at compile time.
If a contract is successfully verified,
dynamic checking of it can be omitted.
This method decreases execution time
without degrading reliable program execution.
In the best case, when all contracts are statically verified,
it provides trust in the software since crashes due to contract
violations cannot occur during program execution.
\\[1ex]
\end{abstract}

\section{Introduction}
\label{sec:intro}

Static types, provided by the programmer or inferred by the compiler,
are useful to detect specific classes of run-time errors at compile time.
This is expressed by Milner \cite{Milner78} as
``well-typed expressions do not go wrong.''
However, not all requirements on operations can be expressed
by standard static type systems.
Hence, one can either refine the type system,
e.g., use a dependently typed programming language
and a more sophisticated programming discipline \cite{Stump16},
or add contracts with pre- and postconditions to operations.
In this paper, we follow the latter approach since it
provides a smooth integration into existing software development
processes. For instance, consider the well-known factorial function:
\begin{curry}
fac n = if n==0 then 1
                else n * fac (n-1)
\end{curry}
Although \code{fac} is intended to work on non-negative natural numbers,
standard static type systems cannot express this constraint so that
\begin{curry}
fac :: Int -> Int
\end{curry}
is provided or inferred as the static type of \code{fac}.\footnote{%
The type inference depends on the underlying static type system.
For instance, Haskell infers a more general overloaded type.}
Although this type avoids the application of \code{fac} on characters
or strings, it allows to apply \code{fac} on negative numbers
which results in an infinite loop.

A \emph{precondition} is a Boolean expression to restrict
the applicability of an operation.
Following the notation proposed in \cite{AntoyHanus12PADL},
a precondition for an operation $f$ is a Boolean operation
with name \code{$f$'pre}. For instance, a precondition
for \code{fac} is
\begin{curry}
fac'pre n = n >= 0
\end{curry}
To use a precondition for checking \code{fac} invocations at run time,
a preprocessor could transform each call to \code{fac}
by attaching an additional test whether the precondition is
satisfied (see \cite{AntoyHanus12PADL}).
After this transformation, an application to \code{fac}
to a negative number results in a run-time error (contract violation)
instead of an infinite loop.

Unfortunately, run-time contract checking requires additional
execution time so that it is often turned off, in particular,
in production systems.
To improve this situation for declarative languages,
we propose to reduce the number of contract checks
by (automatically) verifying them at compile time.
Since we do not expect to verify all of them at compile time,
our approach can be seen as a compromise between
a full static verification, e.g., with proof assistants like Agda, Coq, or
Isabelle, which is time-consuming and difficult,
and a full dynamic checking, which might be inefficient.

For instance, one can verify
(e.g., with an SMT solver \cite{deMouraBjorner08})
that the precondition for the recursive call of \code{fac}
is always satisfied provided that \code{fac} is called
with a satisfied precondition. Hence, we can omit the
precondition checking for recursive calls so that
$n-1$ precondition checks are avoided when we evaluate \code{fac$\;n$}.

In the following, we make this idea more precise
for the functional logic language Curry \cite{Hanus16Curry},
briefly reviewed in the next section,
so that the same ideas can also be applied to purely functional
as well as logic languages.
After discussing contracts for Curry in Sect.~\ref{sec:contracts},
we define a formal model of contract checking
for Curry in Sect.~\ref{sec:contract-verification}.
This is the basis to extract proof obligations for contracts
at compile time.
If these proof obligations can be verified,
the corresponding dynamic checks can be omitted.
Some examples for contract verification are shown
in Sect.~\ref{sec:examples} before we discuss the current
implementation and first benchmark results, which are quite encouraging.

\section{Functional Logic Programming and Curry}
\label{sec:flp}

Functional logic languages combine the most important
features of functional and logic programming in a single language
(see \cite{Hanus13} for a recent survey).
In particular, the functional logic language Curry
\cite{Hanus16Curry}
conceptually extends Haskell with common features of logic programming,
i.e., non-determinism, free variables, and constraint solving.
Since we discuss our methods in the context of functional logic programming,
we briefly review those elements of functional logic languages
and Curry that are necessary to understand the contents of this paper.
More details can be found in surveys on
functional logic programming \cite{Hanus13}
and in the language report \cite{Hanus16Curry}.

The syntax of Curry is close to
Haskell \cite{PeytonJones03Haskell}.
In addition to Haskell, Curry applies rules
with overlapping left-hand sides in a (don't know) non-deterministic manner
(where Haskell always selects the first matching rule)
and allows \emph{free} (\emph{logic}) \emph{variables}
in conditions and right-hand sides of rules.
These variables must be explicitly declared unless they are anonymous.
Function calls can contain free variables, in particular, variables
without a value at call time.
These calls are evaluated lazily where free variables as demanded arguments
are non-deterministically instantiated \cite{AntoyEchahedHanus00JACM}.

\vspace*{-1.0ex}
\begin{example}\label{ex-concdup}\rm
The following simple program shows the functional and logic features
of Curry. It defines an operation \ccode{++} to concatenate two lists,
which is identical to the Haskell encoding.
The operation \code{ins} inserts an element at some (unspecified)
position in a list:
\begin{curry}
(++) :: [a] -> [a] -> [a]      ins :: a -> [a] -> [a]
[]     ++ ys = ys                ins x ys     = x : ys
(x:xs) ++ ys = x : (xs ++ ys)    ins x (y:ys) = y : ins x ys
\end{curry}
\end{example}
Note that \code{ins} is a \emph{non-deterministic operation}
since it might deliver more than one result for a given argument,
e.g., the evaluation of \code{ins$\,$0$\,$[1,2]} yields the values
\code{[0,1,2]}, \code{[1,0,2]}, and \code{[1,2,0]}.
Non-deterministic operations, which are interpreted as
mappings from values into sets of values \cite{GonzalezEtAl99},
are an important feature
of contemporary functional logic languages.
Hence, there is also a predefined \emph{choice} operation:
\begin{curry}
x ? _  =  x
_ ? y  =  y
\end{curry}
Thus, the expression \ccode{0$~$?$~$1} evaluates to \code{0} and \code{1}
with the value non-deterministically chosen.

Non-deterministic operations can be used as any other operation.
For instance, exploiting \code{ins},
we can define an operation \code{perm}
that returns an arbitrary permutation of a list:
\begin{curry}
perm []     = []
perm (x:xs) = ins x (perm xs)
\end{curry}
Non-deterministic operations are quite expressive
since they can be used to completely eliminate logic variables
in functional logic programs.
Actually, it has been shown
that non-deterministic operations and logic variables
have the same expressive power
\cite{AntoyHanus06ICLP,deDiosCastroLopezFraguas07}.
For instance, a Boolean logic variable can be replaced
by the non-deterministic \emph{generator} operation for Booleans
defined by
\begin{curry}
aBool = False ? True
\end{curry}
This equivalence can be exploited when Curry is implemented
by translation into a target language without support for
non-determinism and logic variables.
For instance, KiCS2 \cite{BrasselHanusPeemoellerReck11}
compiles Curry into Haskell by adding a mechanism
to handle non-deterministic computations.
In our case, we exploit this fact by simply ignoring
logic variables since they are considered as syntactic sugar
for non-deterministic value generators.

Curry has many additional features not described here,
like monadic I/O \cite{Wadler97} for declarative input/output,
set functions  \cite{AntoyHanus09} to encapsulate non-deterministic search,
functional patterns \cite{AntoyHanus05LOPSTR} and
default rules \cite{AntoyHanus17TPLP}
to specify complex transformations in a high-level manner,
and a hierarchical module system together with a
package manager\footnote{\url{http://curry-language.org/tools/cpm}}
that provides access to dozens of packages with hundreds of modules.

Due to the complexity of the source language,
compilers or analysis and optimization tools
often use an intermediate language where the
syntactic sugar of the source language has been eliminated
and the pattern matching strategy is explicit.
This intermediate language, called FlatCurry,
has also been used to specify the operational semantics
of Curry programs \cite{AlbertHanusHuchOliverVidal05}.
Since we will use FlatCurry as the basis for verifying contracts,
we sketch the structure of FlatCurry and its semantics.

\begin{figure}[t]
\[
\begin{array}{lcl@{~~~~}l}
P & ::= & D_1 \ldots D_m  & \mbox{(program)} \\
D & ::= & f(x_1,\ldots,x_n) = e  & \mbox{(function definition)} \\
e & ::= & x & \mbox{(variable) } \\
  & | & c(e_1,\ldots,e_n) & \mbox{(constructor call) } \\
  & | & f(e_1,\ldots,e_n)  & \mbox{(function call) } \\
  & | & \mathit{case}~e~\mathit{of}~\{p_1\to e_1; \ldots; p_n \to e_n\}
                         & \mbox{(case expression) } \\
  & | & e_1~\mathit{or}~e_2 & \mbox{(disjunction) } \\
  & | & \mathit{let}~\{x_1 = e_1;\ldots;x_n = e_n\} ~\mathit{in}~ e
       & \mbox{(let binding) } \\
p & ::= & c(x_1,\ldots,x_n)     & \mbox{(pattern)} 
\end{array}
\]
\caption{Syntax of the intermediate language FlatCurry}\label{fig:flatcurry}
\end{figure}

The abstract syntax of FlatCurry is summarized
in Fig.~\ref{fig:flatcurry}.
In contrast to some other presentations
(e.g., \cite{AlbertHanusHuchOliverVidal05,Hanus13}),
we omit the difference between rigid and flexible case expressions
since we do not consider residuation (which becomes less important
in practice and is also omitted in newer implementations
of Curry \cite{BrasselHanusPeemoellerReck11}).
A FlatCurry program consists of a sequence of function definitions,
where each function is defined by a single rule.
Patterns in source programs are compiled into case expressions
and overlapping rules are joined by explicit disjunctions.
For instance, the non-deterministic insert operation \code{ins}
is represented in FlatCurry as
\begin{curry}
$\code{ins}(x,xs) ~=~ (x:xs) ~\mathit{or}~ (\mathit{case}~xs~\mathit{of}~\{ y:ys ~\to~ y : \code{ins}(x,ys) \}$
\end{curry}
The semantics of FlatCurry programs is defined
in \cite{AlbertHanusHuchOliverVidal05}
as an extension of Launchbury's natural semantics for lazy evaluation
\cite{Launchbury93}.
For this purpose, we consider only \emph{normalized} FlatCurry programs,
i.e., programs where the arguments of constructor and function calls 
and the discriminating argument of case expressions
are always variables.
Any FlatCurry program can be normalized by introducing new variables
by let expressions \cite{AlbertHanusHuchOliverVidal05}.
For instance, the expression ``$y : \code{ins}(x,ys)$'' is normalized into
``$\mathit{let}~\{z = \code{ins}(x,ys)\}~\mathit{in}~y : z$.''
In the following, we assume that all FlatCurry programs
are normalized.

In order to model sharing, which is important for lazy evaluation
and also semantically relevant in case of non-deterministic operations
\cite{GonzalezEtAl99}, variables are interpreted
as references into a heap where new let bindings are stored and
function calls are updated with their evaluated results.
To be more precise, a \emph{heap}, denoted by $\Gamma,\Delta,$ or $\Theta$,
is a partial mapping from variables to expressions.
The \emph{empty heap} is denoted by $\emptyheap$.
$\Gamma[x \mapsto e]$ denotes a heap $\Gamma'$ with
$\Gamma'(x) = e$ and $\Gamma'(y) = \Gamma(y)$ for all $x \neq y$.

\begin{figure}[t]
\[
\begin{array}{l@{~~~~~}l}
\rulename{Val} &
  \Gamma : v ~\Downarrow~ \Gamma : v
  ~~~ \mbox{ where $v$ is constructor-rooted } \\[2ex]
\rulename{VarExp} &
  {\displaystyle \frac{\Gamma : e
  ~\Downarrow~
  \Delta : v}{\Gamma[x \mapsto e]:x ~\Downarrow~ \Delta[x\mapsto v] : v} }
\\[4ex]
\rulename{Fun} & {\displaystyle \frac{\Gamma:\rho(e)
~\Downarrow~ \Delta : v} {\Gamma:f(\seq{x_n}) ~\Downarrow~ \Delta :
v}
} ~~ \mbox{ where $f(\seq{y_n})= e \in P$ and $\rho=\{\seq{y_n \mapsto x_n}\}$ } \\[4ex]
\rulename{Let} & {\displaystyle
  \frac{\Gamma[\seq{y_k \mapsto \rho(e_k)}]:\rho(e) ~\Downarrow~ \Delta : v }
{\Gamma: let~ \{\seq{x_k = e_k}\}~ in~e ~\Downarrow~ \Delta : v  }
} ~~~
\begin{array}[c]{l@{ }l}
  \mbox{where }&\rho = \{ \seq{x_k \mapsto y_k} \}\\
  &\mbox{and }\seq{y_k}\mbox{ are fresh variables}
\end{array}\\[4ex]
\rulename{Or} & {\displaystyle  \frac{\Gamma: e_i ~\Downarrow~
\Delta : v } {\Gamma: e_1~or~e_2 ~\Downarrow~ \Delta : v  }
} ~~ \mbox{ where $i \in \{1,2\}$} \\[4ex]
\rulename{Select} &
{\displaystyle
  \frac{\Gamma: x ~\Downarrow~ \Delta : c(\seq{y_n})
         ~~~~~ \Delta : \rho(e_i) ~\Downarrow~ \Theta : v }
       {\Gamma: \mathit{case}~x~\mathit{of}~ \{\seq{p_k \rightarrow e_k}\}
         ~\Downarrow~ \Theta : v  } } ~~
\begin{array}[c]{l@{ }l}
\mbox{ where }& p_i = c(\seq{x_n})\\
&\mbox{ and }\rho = \{ \seq{x_n \mapsto y_n} \}
\end{array}
\end{array}
\]
\caption{Natural semantics of normalized FlatCurry programs}
\label{fig:natsem}
\end{figure}

Using heap structures, one can provide a
high-level description of the operational behavior
of FlatCurry programs in natural semantics style.
The semantics uses judgements of the form
``$\Gamma:e ~\Downarrow~ \Delta:v$'' with the meaning that
in the context of heap $\Gamma$ the expression $e$ evaluates to value
(head normal form) $v$ and produces a modified heap $\Delta$.
Figure~\ref{fig:natsem} shows the rules defining this semantics
w.r.t.\ a given normalized FlatCurry program $P$
($\seq{o_k}$ denotes a sequence of objects $o_1,\ldots,o_k$).

Constructor-rooted expressions (i.e., head normal forms)
are just returned by rule \rulename{Val}.
Rule \rulename{VarExp} retrieves a binding for a variable from the heap
and evaluates it.
In order to avoid the re-evaluation of the same expression,
\rulename{VarExp} updates the heap with the computed value,
which models sharing.
In contrast to the original rules \cite{AlbertHanusHuchOliverVidal05},
\rulename{VarExp} removes the binding from the heap.
On the one hand, this allows the detection of simple loops (``black holes'')
as in functional programming.
On the other hand, it is crucial in combination with non-determinism
to avoid the binding of a variable to different values
in the same derivation (see \cite{Brassel11Thesis} for a detailed
discussion on this issue).
Rule \rulename{Fun} unfolds function calls by evaluating the right-hand side
after binding the formal parameters to the actual ones.
\rulename{Let} introduces new bindings in the heap and renames the
variables in the expressions with the fresh names introduced in the heap.
\rulename{Or} non-deterministically evaluates one of its arguments.
Finally, rule \rulename{Select} deals with $\mathit{case}$ expressions.
When the discriminating argument of $\mathit{case}$ evaluates to a
constructor-rooted term, \rulename{Select} evaluates the corresponding branch
of the $\mathit{case}$ expression.

The FlatCurry representation of Curry programs and its operational semantics
has been used for various language-oriented tools,
like compilers, partial evaluators, or debugging and profiling tools
(see \cite{Hanus13} for references).
We use it in this paper to define a formal model of contract checking
and extract proof obligations for contracts from programs.

\section{Contracts}
\label{sec:contracts}

The use of contracts even in declarative programming languages
has been motivated in Sect.~\ref{sec:intro}.
Contracts in the form of pre- and postconditions as well as
specifications have been introduced into functional logic programming
in \cite{AntoyHanus12PADL}.
Contracts and specifications for some operation are operations
with the same name and a specific suffix.
If $f$ is an operation of type $\tau \to \tau'$,
then a \emph{specification} for $f$ is an operation $\spec{f}$
of type $\tau \to \tau'$,
a \emph{precondition} for $f$ is an operation $\pre{f}$
of type $\tau \to \code{Bool}$, and
a \emph{postcondition} for $f$ is an operation $\post{f}$ of type
$\tau \to \tau' \to \code{Bool}$.

Intuitively, an operation and its specification should be
equivalent operations.
For instance, a specification of non-deterministic list insertion
could be stated with a single rule containing a
functional pattern \cite{AntoyHanus05LOPSTR} as follows:
\begin{curry}
ins'spec :: a -> [a] -> [a]
ins'spec x (xs$\;$++$\;$ys) = xs ++ [x] ++ ys
\end{curry}
A precondition should be satisfied if an operation is invoked,
and a postcondition is a relation between input and output values
which should be satisfied when an operation yields some result.
We have already seen a precondition for the factorial function
in Sect.~\ref{sec:intro}.
A postcondition for the same operation could state that the result
is always positive:
\begin{curry}
fac'post n f = f > 0
\end{curry}
This postcondition ensures the precondition of nested \code{fac} applications,
like in the expression \code{fac$\;$(fac$\;$3)}.
If there is no postcondition but a specification,
the latter can be used as a postcondition.
For instance, a postcondition derived from the specification for \code{ins}
is
\begin{curry}
ins'post :: a -> [a] -> [a] -> Bool
ins'post x ys zs = zs `valueOf` ins'spec$\funset$ x ys
\end{curry}
This postcondition states that the value \code{zs} computed by \code{ins}
is in the set of all values computed by \code{ins'spec}
(where $f\funset$ denotes the set function of $f$, see
\cite{AntoyHanus09}).

Antoy and Hanus \cite{AntoyHanus12PADL} describe a tool which transforms
programs containing contracts and specifications
into programs where these contracts and specifications
are dynamically checked.
This tool is available in recent distributions of the
Curry implementations PAKCS \cite{Hanus16PAKCS}
and KiCS2 \cite{BrasselHanusPeemoellerReck11}
as a preprocessor so that the transformation can be automatically
performed when Curry programs are compiled.
Furthermore, the property-based testing tool CurryCheck \cite{Hanus16LOPSTR}
automatically tests contracts and specifications
with generated input data.

Although these dynamic and static testing tools provide
some confidence in the software under development,
a static \emph{verification} of contracts is preferable since it holds
for all input values, i.e., it is ensured that violations
of verified contracts cannot occur at run time so that
their run-time tests can be omitted.
As a first step towards this objective,
we specify the operational meaning of contract checking
by extending the semantics of Fig.~\ref{fig:natsem}.
Since pre- and postconditions are checked before and after
a function invocation, respectively,
it is sufficient to extend rule \rulename{Fun}.
Assume that function $f$ has a precondition $\pre{f}$ and a
postcondition $\post{f}$ (if some of them is not present,
we assume that they are defined as predicates which always
return \code{True}).
Then we replace rule \rulename{Fun} by the extended rule \rulename{FunCheck}:

\vspace*{-2ex}
\[
 \frac{\Gamma:\pre{f}(\seq{x_n}) ~\Downarrow~ \Gamma' : \code{True} \quad
       \Gamma':\rho(e) ~\Downarrow~ \Delta' : v \quad
       \Delta':\post{f}(\seq{x_n},v) ~\Downarrow~ \Delta : \code{True}}
      {\Gamma:f(\seq{x_n}) ~\Downarrow~ \Delta : v}
\]
where $f(\seq{y_n})= e \in P$ and $\rho=\{\seq{y_n \mapsto x_n}\}$.
For the sake of readability, we omit the normalization of
the postcondition in the premise, which can be added by an
introduction of a $\mathit{let}$ binding for $v$.
The reporting of contract violations can be specified by the following
rules:

\vspace*{-2ex}
\[
 \frac{\Gamma:\pre{f}(\seq{x_n}) ~\Downarrow~ \Gamma' : \code{False}}
      {\Gamma:f(\seq{x_n}) ~\Downarrow~
        \code{<<precondition of }f\code{ violated>>}}
\]
\[
 \frac{\Gamma:\pre{f}(\seq{x_n}) ~\Downarrow~ \Gamma' : \code{True} \quad
       \Gamma':\rho(e) ~\Downarrow~ \Delta' : v \quad
       \Delta':\post{f}(\seq{x_n},v) ~\Downarrow~ \Delta : \code{False}}
      {\Gamma:f(\seq{x_n}) ~\Downarrow~  \code{<<postcondition of }f\code{ violated>>}}
\]
Note that we specified \emph{eager} contract checking,
i.e., pre- and postconditions are immediately and completely
evaluated. Although this is often intended, there are cases
where eager contract checking might influence the execution
behavior of a program, e.g., if the evaluation of a pre- or
postcondition requires to evaluate more than demanded by the original
program. To avoid this problem,
Chitil et al.\ \cite{ChitilMcNeillRunciman04} proposed \emph{lazy}
contract checking where contract arguments are not evaluated
but the checks are performed when the demanded arguments
become evaluated by the application program.
Lazy contract checking could have the problem that the occurrence
of contract violations depend on the demand of evaluation so that
they are detected ``too late.''
Since there seems to be no ideal solution to this problem,
we simply stick to eager contract checking.

\section{Contract Verification}
\label{sec:contract-verification}

In order to statically verify contracts, we have to extract
some proof obligation from the program and contracts.
For instance, consider the factorial function and its precondition,
as shown in Sect.~\ref{sec:intro}.
The normalized FlatCurry representation of the factorial function is
\begin{curry}
fac(n) = let { x = 0 ; y = n==x }
         in case y of True  -> 1
                      False -> let { n1 = n - 1 ; f  = fac n1 }
                                in n * f
\end{curry}
Now consider the call \code{fac($n$)}.
Since we assume that the precondition holds when an operation
is invoked, we know that $n \geq 0$ holds before the case expression
is evaluated.
If the \code{False} branch of the case expression is selected,
we know that $n = 0$ has the value \code{False}.
Altogether, we know that
\[
n \geq 0 \land \lnot (n = 0)
\]
holds when the right-hand side of the \code{False} branch is evaluated.
Since this implies that $n>0$ and, thus, $(n-1) \geq 0$ holds
(in integer arithmetic), we know that the precondition
of the recursive call to \code{fac} always holds.
Hence, its check can be omitted at run time.

This example shows that we have to collect in expressions
(the rules' right-hand sides) properties that are ensured
to be valid when we reach particular points.
For this purpose, we define an
\emph{abstract assertion-collecting semantics}.
It is oriented towards the concrete semantics shown before
but has the following differences:
\begin{enumerate}
\item We compute with symbolic values instead of concrete ones.
\item We collect properties that are known to be valid (also called
\emph{assertions} in the following).
\item Instead of evaluating functions, we collect their pre- and postconditions.
\end{enumerate}

\begin{figure}[t]
\[
\begin{array}{l@{~~~~~}l}
\arulename{Val} &
  \ajudge{\Gamma}{C}{z}{v}{\Gamma}{C \land z = v}
~~~~~
\begin{array}[c]{l@{ }l}
  \mbox{where } & \mbox{$v$ is constructor-rooted or}\\
                & \mbox{$v$ is a variable not bound in $\Gamma$}
\end{array}\\[4ex]

\arulename{VarExp} &
{\displaystyle
 \frac{\ajudge{\Gamma}{C}{z}{e}{\Delta}{D}}
      {\ajudge{\Gamma[x \mapsto e]}{C}{z}{x}{\Delta[x \mapsto e]}{D}}
}\\[4ex]

\arulename{Fun} &
\ajudge{\Gamma}{C}{z}{f(\seq{x_n})}
       {\Gamma}{C \land \pre{f}(\seq{x_n}) \land \post{f}(\seq{x_n},z)} \\[2ex]

\arulename{Let} &
{\displaystyle
 \frac{\ajudge{\Gamma[\seq{y_k \mapsto \rho(e_k)}]}{C}{z}{\rho(e)}{\Delta}{D}}
      {\ajudge{\Gamma}{C}{z}{let~ \{\seq{x_k = e_k}\}~ in~e}{\Delta}{D}}
} ~~~~~
\begin{array}[c]{l@{ }l}
 \mbox{where } & \rho = \{ \seq{x_k \mapsto y_k} \}\\
               & \mbox{and }\seq{y_k}\mbox{ are fresh variables}
\end{array}\\[4ex]

\arulename{Or} &
{\displaystyle
 \frac{\ajudge{\Gamma}{C}{z}{e_1}{\Delta_1}{D_1} \qquad
       \ajudge{\Gamma}{C}{z}{e_2}{\Delta_2}{D_2} }
      {\ajudge{\Gamma}{C}{z}{e_1~or~e_2}{\Delta_1 \cupdot \Delta_2}{D_1 \lor D_2}}
} \\[4ex]

\arulename{Select} &
{\displaystyle
 \frac{\ajudge{\Gamma}{C}{x}{x}{\Delta}{D} \quad
       \ajudge{\Gamma}{D_1}{z}{e_1}{\Theta_1}{E_1} ~\ldots~
       \ajudge{\Gamma}{D_k}{z}{e_k}{\Theta_k}{E_k}
      }
      {\ajudge{\Gamma}{C}{z}{\mathit{case}~x~\mathit{of}~ \{\seq{p_k \to e_k}\}}
       {\Theta_1 \cupdot\ldots\cupdot \Theta_k}{E_1 \lor \ldots \lor E_k} } } \\[2ex]
 & \mbox{where $D_i = D \land x=p_i$ ($i=1,\ldots,k$)}
\end{array}
\]
\caption{Abstract assertion-collecting semantics}
\label{fig:abssem}
\end{figure}

\noindent
The abstract semantics uses judgements of the form
``$\ajudge{\Gamma}{C}{z}{e}{\Delta}{D}$'' where
$\Gamma$ is a heap,
$z$ is a (result) variable, $e$ is an expression,
and $C$ and $D$ are \emph{assertions}, i.e., Boolean formulas
over the program signature.
Intuitively, this judgement means that if $e$ is evaluated to $z$
in the context $\Gamma$ where $C$ holds,
then $D$ holds after the evaluation.

Figure~\ref{fig:abssem} shows the rules defining this
abstract semantics.
Rule \arulename{Val} immediately returns the collected assertions.
Since this semantics is intended to compute with symbolic values,
there might be variables without a binding to a concrete value.
Hence, \arulename{Val} also returns such unbound variables.
Rule \arulename{VarExp} behaves similarly to rule \rulename{VarExp}
of the concrete semantics and returns the assertions collected
during the abstract evaluation of the expression.
Note that the abstract semantics does not really evaluate
expressions since it should always return the collected assertions
in a finite amount of time.
For the same reason,
rule \arulename{Fun} does not invoke the function in order
to evaluate its right-hand side. Instead, the pre- and postcondition
information is added to the collected assertions since they
must hold if the function returns some value.
The notation $\pre{f}(\seq{x_n})$ and $\post{f}(\seq{x_n},z)$
 in the assertion means
that the logical formulas corresponding to the pre- and postcondition
are added as an assertion.
These formulas might be simplified by replacing occurrences
of operations defined in the program by their definitions.
Rule \arulename{Let} adds the let bindings to the heap,
similarly to the concrete semantics, before evaluating the argument
expression.
Rules \arulename{Or} and \arulename{Select} collect
all information derived from alternative computations,
instead of the non-deterministic concrete semantics.
Rule \arulename{Select} also collects inside each branch
the condition that must hold
in the selected branch, which is important to get
precise proof obligations.
To avoid the renaming of local variables in different branches,
we implicitly assume that all local variables are unique in a normalized
function definition.

In contrast to the concrete semantics, the abstract semantics
is deterministic, i.e., for each heap $\Gamma$, assertion $C$,
variable $z$, and expression $e$, there is a unique (up to variable
renamings in let bindings) proof tree and assertion $D$
so that the judgement ``$\ajudge{\Gamma}{C}{z}{e}{\Delta}{D}$''
is derivable.

The abstract semantics allows to extract proof obligations
to verify contracts.
For instance, to verify that a postcondition $\post{f}$
for some function $f$ defined by $f(\seq{x_n}) = e$ holds,
one derives a judgement
(where $z$ is a new variable)
\[
\ajudge{\emptyheap}{\pre{f}(\seq{x_n})}{z}{e}{\Gamma}{C}
\]
and proves that $C$ implies $\post{f}(\seq{x_n},z)$.

As an example, consider the non-deterministic operation
\begin{curry}
coin = 1 $\mathit{or}$ 2
\end{curry}
and its postcondition
\begin{curry}
coin'post z = z > 0
\end{curry}
(the precondition is simply \code{True}).
We derive for the right-hand side of \code{coin}
the following proof tree:
\begin{prooftree}
\AxiomC{}
\LeftLabel{\arulename{Val}}
\UnaryInfC{$\ajudge{\emptyheap}{true}{z}{1}{\emptyheap}{z=1}$}
\AxiomC{}
\RightLabel{\arulename{Val}}
\UnaryInfC{$\ajudge{\emptyheap}{true}{z}{2}{\emptyheap}{z=2}$}
\LeftLabel{\arulename{Or}}
\BinaryInfC{$\ajudge{\emptyheap}{true}{z}{1~\mathit{or}~2}{\emptyheap}{z=1 \lor z=2}$}
\end{prooftree}
Since $z=1 \lor z=2$ implies $z>0$, the postcondition of \code{coin}
is always satisfied.

If we construct the proof tree for the right-hand side $e$ of
the factorial function, we derive the following judgement:
\[
\acjudge{\emptyheap}{n \geq 0}{z}{e}
       {(n \geq 0 \land y=true \land z=1) \lor (n \geq 0 \land y=false)}
\]
Since there is no condition on the result variable $z$ in the
second argument of the disjunction,
this assertion does not imply the postcondition $z > 0$.
The reason is that the recursive call to \code{fac}
is not considered in the proof tree since it does not occur
at the top level. Note that rule \arulename{Fun} only adds
the contract information of top-level operations but no contracts
of operations occurring in arguments.
Due to the lazy evaluation strategy,
one does not know at compile time whether some argument expression
is evaluated. Hence, it would not be correct to add the contract
information of nested arguments.
For instance, consider the operations
\begin{curry}
const x y = y       f x | x>0 = 0          g x = const (f x) 42
                    f'post x z = x>0
\end{curry}
If $e$ denotes the right-hand side of \code{g} (in normalized FlatCurry form),
then we can derive with the inference rules of Fig.~\ref{fig:abssem}
the judgement
\[
\ajudge{\emptyheap}{true}{z}{e}{[y1 \mapsto \code{f}(x), y2 \mapsto 42]}{true}
\]
If we change rule \arulename{Fun} so that the contracts
of argument calls are also added to the returned assertion,
then we could derive
\[
\ajudge{\emptyheap}{true}{z}{e}{[y1 \mapsto \code{f}(x), y2 \mapsto 42]}{x>0}
\]
This postcondition is clearly wrong since \code{(g$\;$0)} successfully
evaluates to \code{42}.

Nevertheless, we can improve our abstract semantics
in cases where it is ensured that arguments are evaluated.
For instance, primitive operations, like \code{+}, \code{*}, or \code{==},
evaluate their arguments. Thus, we can add the following rule
(and restrict rule \arulename{Fun} to exclude these operations):
\[
\begin{array}{l@{~~}l}
\arulename{PrimOp} &
{\displaystyle
 \frac{\ajudge{\Gamma}{C}{x}{x}{\Delta}{D} \quad
       \ajudge{\Gamma}{D}{y}{y}{\Theta}{E}}
      {\ajudge{\Gamma}{C}{z}{x~\oplus~y}{\Theta}{E \land z = x~\oplus~y}} }
 ~~\mbox{where } \oplus \in \{\code{==}, \code{+},\code{-},\code{*},\ldots\}
\end{array}
\]
Since primitive operations are often known to the underlying verifier,
we also collect the information about the call of the primitive operation.
In a similar way, one can also improve user-defined functions
if some argument is known to be demanded,
a property which can be approximated at compile time
by a demand analysis \cite{Hanus12ICLP}.

If we construct a proof tree for the factorial function
with these refined inference rules, we obtain the
following (simplified) assertion:
\[
(n \geq 0 \land n=0 \land z=1) \lor
(n \geq 0 \land n \neq 0 \land n1 \geq 0 \land f>0 \land z=n*f)
\]
Since this assertion implies $z > 0$, the postcondition $\post{\code{fac}}$
holds so that its checking can be omitted at run time.

Proof obligations for preconditions can also be extracted
from the proof tree.
For this purpose, one has to consider occurrences of
operations with non-trivial preconditions.
If such an operation occurs as a top-level expression or in a let binding
associated to a top-level expression and the assertion
before this expression implies the precondition,
then one can omit the precondition checking for this call.
For instance, consider again the proof tree for the
right-hand side of the factorial function
which contains the following (simplified) judgement:
\[
\ajudge{\emptyheap}{n \geq 0 \land n \neq 0}
 {z}{\mathit{let}~\{n1=n-1;f = fac~n1\}~\mathit{in}~n*f}{\ldots}{\ldots}
\]
Since $n \geq 0 \land n \neq 0 \land n1 = n-1$ implies $n1 \geq 0$,
the precondition holds so that its check can be omitted for this
recursive call.

The correctness of our approach relies on the following
relation between the concrete and the abstract semantics:
\begin{theorem}
If $\Gamma: e ~\Downarrow~ \Gamma' : v$ is a valid judgement,
$z$ a variable, and $C$ an assertion such that $\widehat{\Gamma} \Rightarrow C$
is valid,
then there is a valid judgement
$\ajudge{\Gamma}{C}{z}{e}{}{D}$
with $(\widehat{\Gamma'} \land z=v) \Rightarrow D$.
\end{theorem}
Here, $\widehat{\Gamma}$ denotes the representation of
heap information as a logic formula, i.e.,
\[
\widehat{\Gamma} = \bigwedge \{ x=e \mid x \mapsto e \in \Gamma,~ e \mbox{ not operation-rooted} \}
\]
The proof is by induction on the height of the proof tree
and requires some technical lemmas which we omit here
due to lack of space.

\section{More Examples}
\label{sec:examples}

There are various recursively defined operations
with pre- and postconditions that can be verified
similarly to \code{fac} as shown above.
For instance, the postcondition and the preconditions
for both recursive calls to \code{fib}
in
\begin{curry}
fib x | x == 0    = 0
      | x == 1    = 1
      | otherwise = fib (x-1) + fib (x-2)$\listline$
fib'pre n = n >= 0
fib'post n f = f >= 0
\end{curry}
can be verified with a similar reasoning.
The precondition on \code{take} defined by
\begin{curry}
take 0 xs           = []
take n (x:xs) | n>0 = x : take (n-1) xs$\listline$
take'pre n xs = n >= 0
\end{curry}
can be similarly verified since the list structures are not relevant here.
On the other hand, the verification of the precondition of
the recursive call of the function \code{last} defined by
\begin{curry}
last [x]      = x
last (_:x:xs) = last (x:xs)$\listline$
last'pre xs = not (null xs)
\end{curry}
requires the verification of the implication
\[
not\;(null\;xs) \land xs=(y{:}ys) \land ys=(z{:}zs) \Rightarrow not\;(null\;(z{:}zs))
\]
This can be proved by evaluating the right-hand side to $true$.
Hence, a reasonable verification strategy includes the simplication
of proof obligations by symbolic evaluation
before passing them to the external verifier.\footnote{%
Since Curry programs might contain non-terminating operations,
one has to be careful when simplifying expressions.
In order to ensure the termination of the simplification process,
one can either limit the number of simplification steps
or use only operations for simplification that are known to
be terminating. Since the latter property can be approximated
by various program analysis techniques,
the Curry program analyzer CASS \cite{HanusSkrlac14}
contains such an analysis.}

A more involved operation is the list index operator
which selects the $n$th element of a list:
\begin{curry}
nth (x:xs) n | n==0 = x
             | n>0  = nth xs (n-1)$\listline$
nth'pre xs n = n >= 0 && length (take (n+1) xs) == n+1
\end{curry}
The precondition ensures that the element to be selected always exists
since the selected position is not negative and not larger
than the length of the list. The use of the operation \code{take}
(instead of the simpler condition \code{length$\;$xs > n})
is important to allow the application of \code{nth} also to infinite lists.
To verify that the precondition holds for the recursive call,
one has to verify that
\[
n \geq 0 \land length\;(take\;(n+1)\;xs) = n+1 \land xs=(y{:}ys) \land n \neq 0 \land n>0
\]
implies
\[
(n-1) \geq 0 ~\land~ length\;(take\;((n-1)+1)\;ys) = (n-1)+1
\]
The proof of the first conjunct uses reasoning on integer arithmetic
as in the previous examples.
The second conjunct can also be proved by SMT solvers
when the rules of the operations \code{length} and \code{take}
are axiomatized as logic formulas.

\section{Implementation and Benchmarks}
\label{sec:impl}

We have implemented static contract verification
as a fully automatic tool which
tries to verify contracts at compile time and, in case of a
successful verification,
removes their run-time checking from the generated code.
The complete compilation chain with this tool is as follows:
\begin{enumerate}
\item
The Curry preprocessor performs a source-level transformation
to add contracts as run-time checks,
as sketched in Sect.~\ref{sec:contracts} and described in
\cite{AntoyHanus12PADL}.
\item
The preprocessed program is compiled with the standard Curry front end
into an intermediate FlatCurry program.
\item
For each contract, the contract verifier 
extracts the proof obligation as described in
Sect.~\ref{sec:contract-verification}.
\item
Each proof obligation is translated into SMT-LIB format
and sent to an SMT solver (here: Z3 \cite{deMouraBjorner08}).
\item
If the proof shows the validity of the contract,
its check is removed from the FlatCurry program.
\end{enumerate}
This general approach can be refined.
For instance, if a pre- or postcondition is a conjunction of formulas,
each conjunct can separately be verified and possibly removed.
This allows to make dynamic contract checking more efficient
even if the complete contract cannot be verified.

Although our tool is a prototype,
we applied it to some initial benchmarks
in order to get an idea about the efficiency improvement
by static contract verification.
For this purpose, we compared the execution time of the program
with and without static contract checking.
Note that in case of preconditions, only verified
preconditions for recursive calls can be omitted
so that the operations can safely be invoked as before.

For the benchmarks, we used the Curry implementation
KiCS2 (Version 0.6.0) \cite{BrasselHanusPeemoellerReck11}
with the Glasgow Haskell Compiler (GHC 7.10.3, option -O2) as its back end
on a Linux machine (Debian 8.9) 
with an Intel Core i7-4790 (3.60Ghz) processor and 8GiB of memory.
Table~\ref{table:benchmarks} shows the execution times
(in seconds, where ``0.00'' means less than 10 ms) of executing
a program with the given main expression.
Column ``dynamic'' denotes purely dynamic contract checking
and column ``static+dynamic'' denotes
the combination of static and dynamic contract checking as
described in this paper.
The column ``speedup'' is the ratio of the previous columns
(where a lower bound is given if the execution time of the optimized
program is below 10 ms).

\begin{table}[t]
\begin{center}
\begin{tabular}{|l|c|c|c|}
\hline
Expression & dynamic & static+dynamic & speedup \\
\hline
\code{fac 20} & 0.00 & 0.00 & n.a. \\
\code{sum 1000000} & 0.84 & 0.22 & 3.88 \\
\code{fib 35} & 1.95 & 0.60 & 3.23 \\
\code{last [1..20000000]} & 0.63 & 0.35 & 1.78 \\
\code{take 200000 [1..]} & 0.31 & 0.19 & 1.68 \\
\code{nth [1..] 50000} & 26.33 & 0.01 & 2633 \\
\code{allNats 200000} & 0.27 & 0.19 & 1.40 \\
\code{init [1..10000]} & 2.78 & 0.00 & $>$277 \\
\code{[1..20000] ++ [1..1000]} & 4.21 & 0.00 & $>$420 \\
\code{nrev [1..1000]} & 3.50 & 0.00 & $>$349 \\
\code{rev [1..10000]} & 1.88 & 0.00 & $>$188 \\
\hline
\end{tabular}
\end{center}
\caption{Benchmarks comparing dynamic and static contract checking}
\label{table:benchmarks}
\end{table}

Many of the programs that we tested are already discussed in this paper.
\code{sum} is similar to \code{fac} but adds all numbers instead of
multiplying them.
\code{allNats} produces (non-deterministically) some natural number
between \code{0} and the given argument, where the precondition
requires that the argument must be non-negative.
\code{init} removes the last element of a list,
where the precondition requires that the list is non-empty and
the postcondition states that the length of the output list
is decremented by one.
The list concatenation (\code{++}) has a postcondition
which states that the length of the output list
is the sum of the lengths of the input lists.
\code{nrev} and \code{rev} are naive and linear list reverse
operations, respectively, where their postconditions require
that the input and output lists are of identical length.

As expected, the benchmarks show that static contract checking has
a positive impact on the execution time.
If contracts are complex, e.g., require recursive computations
on arguments, as in \code{nth}, \code{init}, \ccode{++}, or \code{rev},
static contract checking can improve the execution times
by orders of magnitudes.
Even if the improvement is small or not measurable (e.g., \code{fac}),
static contract verification is useful since any verified contract
increases the confidence in the correctness of the software and
contributes to a more reliable software product.

\section{Related Work}
\label{sec:related}

As contract checking is an important contribution
to obtain more reliable software,
techniques for it have been extensively explored.
Mostly related to our approach is the work of
Stulova et al.\ \cite{StulovaMoralesHermenegildo16}
on reducing run-time checks of assertions by static analysis
in logic programs.
Although the objectives of this and our work are similar,
the techniques and underlying programming languages
are different.
For instance, Curry with its demand-driven evaluation strategy
prevents the construction of static call graphs that
are often used to analyze the data flow as in logic programming.
The latter is used by Stulova et al.\ where
assertions are verified by static analysis methods.
Hence, the extensive set of benchmarks presented in their work
is related to typical abstract domains
used in logic programming, like modes or regular types.
There are also approaches to approximate argument/result size relations
in logic programs, e.g., \cite{SerranoEtAl13},
which might be used to verify assertions related to the size of data.
On the other hand, many of our examples require
symbolic reasoning on integer arithmetic with user-defined functions.
For this purpose, SMT solvers are well suited and
we have shown that they can be successfully applied to
verify complex assertions (see example \code{nth} above).

Static contract checking has also been explored
in purely functional languages.
For instance, \cite{XuPeytonJonesClaessen09} presents a method
for static contract checking in Haskell by a program transformation
and symbolic execution. Since an external verifier is not used,
the approach is more limited.
Another approach is the extension of the type system
to express contracts as specific types.
Dependent types are quite powerful since they allow
to express size or shape constraints on data in the language of types.
Although this supports the development of programs together with their
correctness proofs \cite{Stump16}, programming in such a language
could be challenging if the proofs are difficult to construct.
Therefore, we prefer a more practical method by checking
properties which cannot be statically proved at run time.
Another approach to express contracts as types is
LiquidHaskell \cite{VazouSeidelJhala14,VazouSeidelJhalaPeytonJones14}.
Similarly to our approach, LiquidHaskell uses an external SMT solver
to verify contracts.
Hence, LiquidHaskell can verify quite complex assertions,
as shown by various case studies in \cite{VazouSeidelJhala14}.
Nevertheless, there might be assertions that cannot be verified
in this way so that a combination of static and dynamic checking
is preferable in practice.

An alternative approach to make dynamic contract checking
more efficient has been proposed in \cite{DimoulasPucellaFelleisen09}
where assertions are checked in parallel to the
application program.
Thus, one can exploit the power of multi-core computers
for assertion checking by running the main program
and the contract checker on different cores.

\section{Conclusions}
\label{sec:conclusions}

In this paper we proposed a framework to combine static and
dynamic contract checking.
Contracts are useful to make software more reliable, e.g.,
avoid invoking operations with unintended arguments.
Since checking all contracts at run time increases the overall
execution time, we have shown a method to verify contracts in Curry
at compile time by using an external SMT solver.
Of course, this might not be successful in all cases
so that unverified contracts are still required to be checked at run time.
Nevertheless, our initial experiments show the advantages
of this technique, in particular, to reduce dynamic contract
checking for recursive calls.
Since we developed this framework for Curry, a language
combining functional and logic programming features,
the same techniques can be applied to purely functional or
purely logic languages.

We do not expect that all contracts can be statically verified.
Apart from the complexity of some contracts,
preconditions of operations of the API of some libraries or packages
cannot be checked since their use is unknown at compile time.
However, one could provide two versions of such operations,
one with a dynamic precondition check and one (``unsafe'') without
this check.
Whenever one can verify that the precondition is satisfied at the call site,
one can invoke the version without the precondition check.
If all versions with precondition checks become dead code in
a complete application, one has a high confidence in the quality
of the entire application.

For future work, we will improve our tool in order to test
the effectiveness of our approach on larger examples.
This might provide also insights how to improve this approach
in practice, e.g., how to use demand information to generate
more precise proof obligations.
If the contract verifier finds counter-examples to some proof obligation,
one could also analyze these in order to check whether they show
an actual contract violation.
Furthermore, it might also be interesting to improve the power
of static contract checking by integrating
abstract interpretation techniques,
like \cite{FaehndrichLogozzo11,StulovaMoralesHermenegildo16}.

\paragraph{Acknowledgments.}
The author is grateful to the anonymous reviewers
for their helpful comments.

\end{document}